\documentclass[preprint,10pt]{aastex}

\usepackage{graphicx} 
\usepackage{multicol}
\usepackage{natbib}
\usepackage{wrapfig}

\newcommand{\be}{\begin{equation}}
\newcommand{\ee}{\end{equation}}
\newcommand{\nn}{\mbox{} \nonumber \\ \mbox{} }
\newcommand{\ba}{\begin{eqnarray}}
\newcommand{\ea}{\end{eqnarray}}
\newcommand{\om}{\omega}
\newcommand{\Alfven}{ Alfv\'{e}n }

\newcommand{\E}{{\bf E}}
\newcommand{\B}{{\bf B}}

\renewcommand{\v}{{\bf v}}

\newcommand{\Fermi}{{\it Fermi}}

\newcommand\etal{\textit{et al.\ }}
\newcommand\eg{\textit{e.g.\ }}

\newcommand\lo{\mathrel{\raise.3ex\hbox{$<$}\mkern-14mu\lower0.6ex\hbox{$\sim$}}}
\newcommand\go{\mathrel{\raise.3ex\hbox{$>$}\mkern-14mu\lower0.6ex\hbox{$\sim$}}}

\newcommand{\Bf}{{magnetic field}}
\newcommand{\Bfs}{{magnetic fields}}
\newcommand{\Ef}{{electric  field}}
\newcommand{\Efs}{{electric fields}}
\newcommand{\CR}{{cosmic ray}}
\newcommand{\CRs}{{cosmic rays}}
\newcommand{\NS}{neutron star}
\newcommand{\NSs}{{neutron stars}}
\newcommand{\EM}{electromagnetic}

\newcommand{\Sc}{Schwarzschild}
\newcommand{\ms}{magnetosphere}
\newcommand{\mss}{magnetospheres}

\begin{document}

\author{Maxim Lyutikov\\
Department of Physics, Purdue University, 
 525 Northwestern Avenue,
West Lafayette, IN
47907-2036 USA\\ INAF - Osservatorio Astrofisico di Arcetri
Largo Enrico Fermi 5, I - 50125 Firenze. Italia \\
The Canadian Institute for Theoretical Astrophysics,
University of Toronto, 
60 St. George Street 
Toronto, Ontario, M5S 3H8
Canada}

\author{A. Lazarian\\
Department of Astronomy, University of Wisconsin-Madison,
475 Charter St., Madison, WI 53706}

\title{Topics in microphysics of relativistic plasmas}

\begin{abstract}
Astrophysical plasmas can have parameters vastly different from the more studied laboratory and space plasmas. In particular, the magnetic fields can be the dominant component of the plasma, with energy-density exceeding the particle rest-mass energy density. Magnetic fields  then determine the plasma dynamical evolution, energy dissipation and acceleration of non-thermal particles. Recent data coming from astrophysical high energy missions, like magnetar bursts and  Crab nebula flares, point to the importance of magnetic reconnection in these objects.

 In this review we outline a broad spectrum of problems related to the astrophysical relevant processes in magnetically dominated relativistic plasmas. We discuss the problems of large scale dynamics of relativistic plasmas, relativistic reconnection and particle acceleration at reconnecting layers, turbulent cascade in force-fee plasmas. A number of astrophysical applications are also discussed.
 \end{abstract}

\section{Introduction}

In many astrophysical settings the magnetic field controls the overall dynamics
of plasma while  the dissipation of magnetic energy may power the high energy
emission.  
The relevant  astrophysical settings include magnetars (strongly magnetized \NSs possessing
super-strong \Bfs), pulsars and pulsar wind nebulae, jets of Active Galactic
Nuclei and Gamma-Ray Bursters.  All these objects are efficient emitters of X-rays
and $\gamma$-rays and in the past two decades they have been subjects of 
intensive observational studies using a number of very successful high energy
satellites.  These objects seem to 
share one important property -- their plasma is magnetically dominated, that is,  
the energy density of this plasma is dominated not by the rest mass-energy of 
matter but by the mass-energy of magnetic field. This is dramatically different  
from the laboratory plasmas, the magnetospheres of planets, and the interplanetary 
plasma. 

Recently, these topics came to the front of astrophysical and plasma physical research, driven by a series of highly successful high energy mission like, {\it Swift},  \Fermi, AGILE satellites and coming on-line of the very high energy $\gamma$-ray telescopes like HESS and VERITAS.
{\it A number of observations point to the importance of magnetic dissipation in astrophysical high energy sources,} as we describe below. This  signifies a shift of paradigm (from the fluid-dominated point of view) and requires a targeted study of plasma microphysics in a new regime.  In this review we outline the related basic plasma physical problems and possible astrophysical applications.

\subsection{Crab nebula flares:  a new type of astrophysical events} The constancy of the high energy Crab nebula emission has been surprisingly shown to be false by multiple day- to week-long flares, presenting a challenge to standard pulsar wind models \citep{kc84}.  During these events, the Crab Nebula gamma-ray flux above 100 MeV exceeded its average value by a factor of several or higher \citep{Abdo:2011,Tavani:2011,Buehler:2012}, while in other energy bands nothing unusual was observed \citep[e.g.][and references therein]{Abdo:2011,Tavani:2011}.  Additionally, sub-flare variability timescales of $\sim 10$ hours has been observed \citep{Buehler:2012}.
The prevailing conclusion from the observations of flares is that 
flares are associated with the nebular (and  not the \NS)   and are mostly likely due to the highest energy synchrotron emitting electrons. Thus,  the flares reflect the instantaneous injection/emission properties of  the nebular and are not expected to produce a noticeable change in the inverse Compton (IC) component above $\sim 1 $ GeV.  One of the most surprising property of the flares is their short time-scale variability, with typical duration two orders of magnitude smaller than the dynamical time-scale of the nebular. 

These events question the dominant paradigm of shock acceleration in pulsar wind nebular \citep{2010MNRAS.405.1809L,2012MNRAS.426.1374C}. The key argument in favor of the reconnection origin of the flare is its SED: the peak frequency is above the classical synchrotron limit \citep{1996ApJ...457..253D,2010MNRAS.405.1809L}.  This limit comes from assuming the electric field accelerating the emitting particles, $E$, is less than the emission region magnetic field, or or $E=\eta B$, where $0<\eta<1$.  
\be
  {\cal E}_{\rm ph}^{\rm max} =   \frac{27}{16\pi}\, \eta\, \frac{mhc^3}{e^2} = 236\,  \eta\, \mbox{MeV}\, ,
   \ee
    Instead, these events  offer tantalizing evidence in favor of relativistic reconnection \citep{begelman_11}.
 \cite{2012MNRAS.426.1374C} suggested that the flare arises due to intermitted reconnection  in the downstream region, which produces relativistically moving blobs of plasma.   
 We associate the duration of the flare with stochastically changing properties of plasma within the nebula.  Second, the flares are apparently isolated, intermittent high flux events.  Such intermittent behavior is often associated with power-law distributions of various kinds generated by astrophysical systems such as magnetically-driven Solar flares  \citep[][and references therein]{2005psci.book.....A}.

Relativistic reconnection is a natural flaring mechanism in PWNe.  The flare can be due to a highly localized emission region, or blob, so that the flare observables determine the intrinsic properties of the emission region.  The natural flaring mechanism in this category is relativistic magnetic reconnection, which has been invoked by Crab Nebula flare models \citep{2011ApJ...737L..40U} and fast flaring models in gamma-ray bursts (GRBs) and active galactic nuclei \citep[AGN,][]{LyutikovRem,2009MNRAS.395L..29G}.  

 \begin{figure}[h!]
\includegraphics[width=0.99\textwidth]{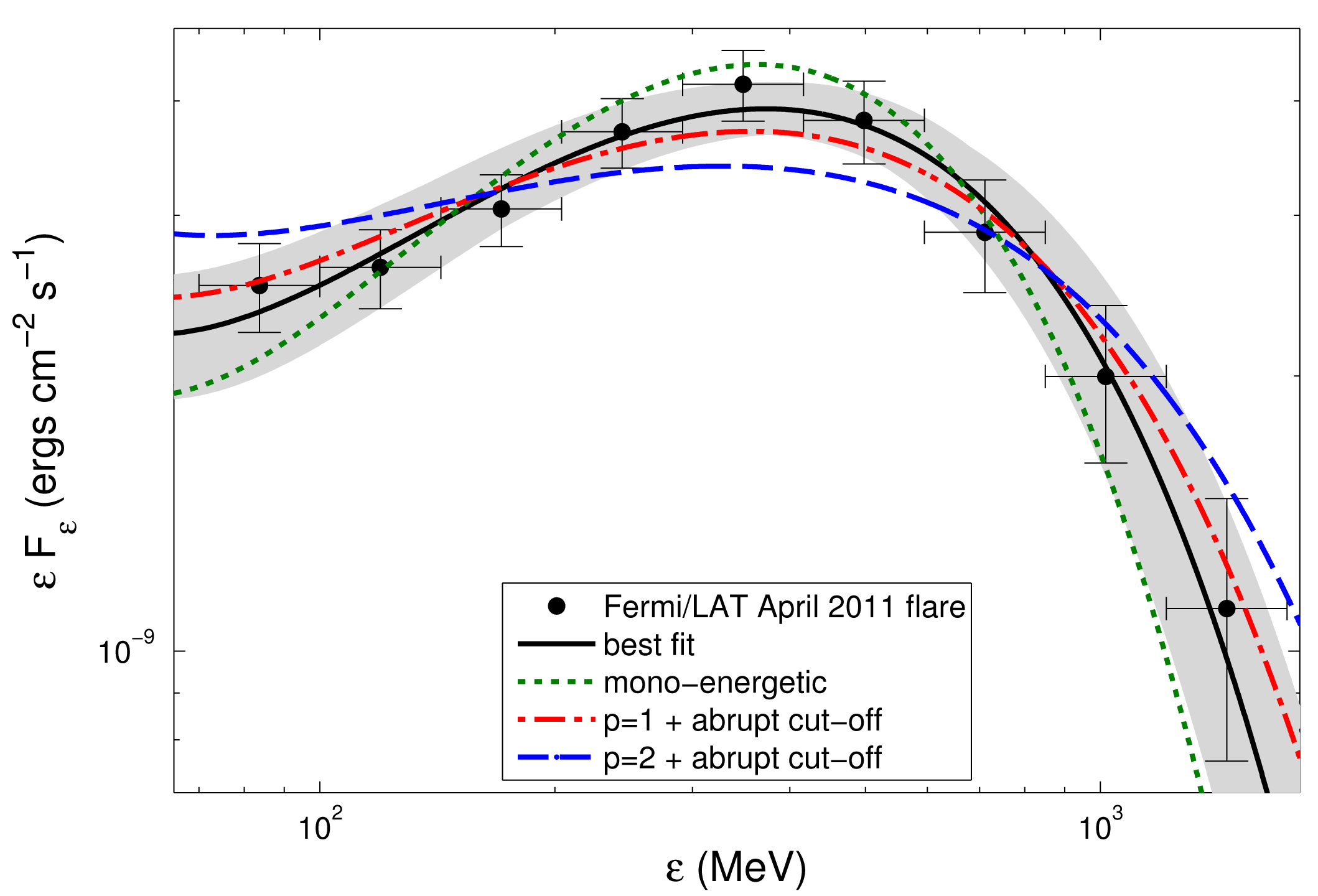}
\caption{ The Fermi/LAT data from the most energetic part of the April 2011 Crab  flare  \protect\citep[][]{Buehler:2012} with the corresponding best fit curve  and SEDs from three different electron energy distributions: a $p=1$ power-law with an abrupt cut-off, the same for $p=2$, and a mono-energetic electron distribution   \protect\citep[][]{2012MNRAS.426.1374C} .  The shaded area represents the one-$\sigma$ error region. The data favor steep injection spectrum with a pile-up, consistent with acceleration in reconnection regions.}
\label{bestfit} 
  \end {figure}

 In reconnection,  the magnetic energy of a localized region, a current sheet, is converted to random particle energy, (possibly) bulk relativistic motion, and radiation \citep[for studies on reconnection in highly magnetized relativistic plasmas, see][]{LyutikovUzdensky,2005MNRAS.358..113L,2011ApJ...737L..40U,mu10}.  Reconnection in PWNe has already been studied in the past as a possible resolution of the well known $\sigma$-problem \citep{LyubarskyStripedWind,2012arXiv1207.3192K}.  In a similar vein, \cite{2010MNRAS.405.1809L} proposes a model in which, reconnection occurs primarily along the rotation axis and equatorial region well beyond the light cylinder, thus qualitatively reproducing the jet/equatorial wisp morphology of the nebula.

\begin{figure}[h!]
\centering
\includegraphics[width=0.99\linewidth]{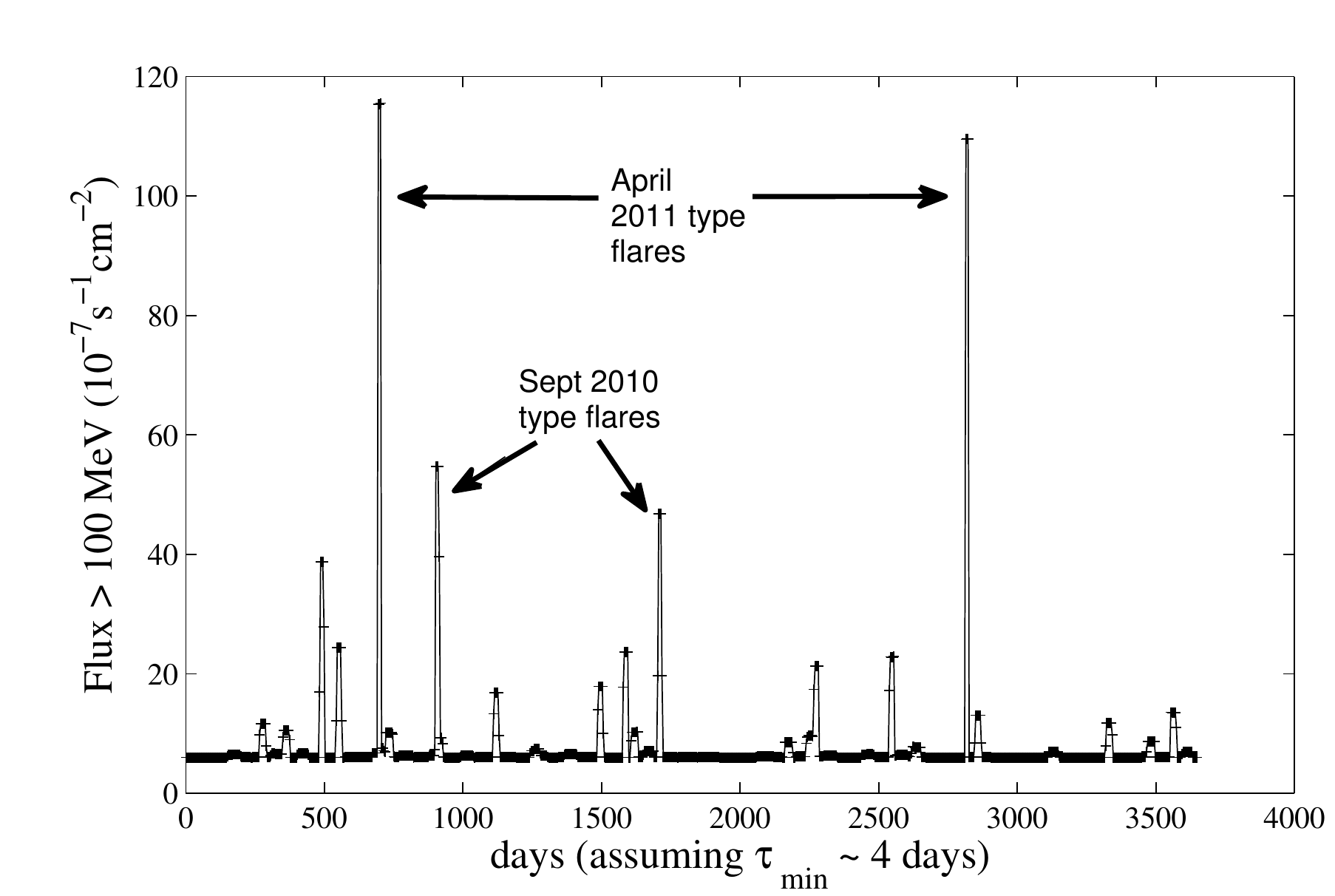}
\caption{Ten year simulated Crab nebula light curve with the reconnection model of Crab flares.  The ``April 2011 type flares'' represent flares with increases of $\sim$ 30 over the nebular average as found in \protect\cite{Buehler:2012}. } 
\label{power}
\end{figure}

We  have  developed \citep{2012MNRAS.426.1374C} a statistical model of the emission from Doppler boosted reconnection mini-jets, looking for analytical expressions for the moments of the resulting nebula light curve (e.g. time average, variance, skewness). 
The light curve has a flat power spectrum that transitions at short timescales to a decreasing power-law of index 2, Fig. \ref{power}. The flux distribution from mini-jets follows a decreasing power-law of index $\sim 1$, implying the average flux from flares is dominated by bright rare events. The predictions for the flares' statistics can be tested against forthcoming observations.  We find the observed flare spectral energy distributions (SEDs) have several notable features: A hard power-law index of $p\lesssim 1$ for accelerated particles that is expected in various reconnection models, including some evidence of a pile-up near the radiation reaction limit. Also, the photon energy at which the SED peaks is higher than that implied by the synchrotron radiation reaction limit, indicating the flare emission regions' Doppler factors are $\gtrsim$ few.  Magnetic reconnection can be an important, if not dominant, mechanism of particle acceleration within the nebula.  

If magnetic reconnection is what causes the flares, the reconnection process itself may leave a particular imprint on the SED in the form of a hard electron distribution.  If much of the synchrotron emission occurs from particles near the synchrotron limit, then the emitting particles will display a SED that is close the single-particle synchrotron SED, see Fig. \ref{bestfit}. 
 The best flare SED observations to date are 11 SEDs taken during the April 2011 by Fermi/LAT team \citep{Buehler:2012}.  They were fitted with an empirical function of the form $\epsilon F_{\epsilon}\propto \epsilon^{a}\exp{(\epsilon/\epsilon_c)}$, where $\epsilon$ represents photon energy, and different values for the normalization and $\epsilon_c$ were used for each SED.  The parameter $a$ was assumed to be constant for all of the SEDs, and its best fit value is $a=0.73 \pm 0.12$.  The SED taken during the most luminous part of the flare, with a $\epsilon F_{\epsilon}$ maximum of $\sim 4\times10^{-9}$ ergs$^{-1}$ s$^{-1}$ at a peak photon energy of $\epsilon_{\rm peak}=375 \pm 26$ MeV, probably constrained the best fit value of $a$ the most. 
  
   \subsection{Magnetar Giant Flares.}
 In phenomena  possibly related to Crab flares, two closely related classes of young neutron stars -- Anomalous X-ray Pulsars
(AXPs) and the Soft Gamma-ray Repeaters (SGRs) -- both show X-ray flares and,
once localized, quiescent X-ray emission (Kouveliotou et al. 1998, Gavriil
\etal\ 2002; for recent review see \cite{wt04}). Their high energy emission is
powered by the   dissipation of their super-strong \Bfs, $B > 10^{15}$G (Thompson \&
Duncan 1996). Two models of GFs are proposed. First, a GF may result from a {\it  sudden 
untwisting of the internal magnetic
field}  \citep[and twisting-up of the external \Bf,][]{td95,thompson01}.  Alternatively, a {\it slow}  untwisting  of the internal magnetic field may
lead  to a gradual twisting of  magnetospheric  field lines, on time scales much longer than  the   GF, until 
it reaches a dynamical stability threshold  due to increasing energy associated with the
current-carrying magnetic field. Then follows a 
{\it sudden relaxation} of the  twist outside the
star and  an associated dissipation and a change of  magnetic topology
lead to a GF,  in  analogy with solar flares and Coronal Mass Ejections (CMEs)  \citep{lyutikov06}.

 The observed
sharp rise of $\gamma$-ray flux during GF, on a time scale similar
to the \Alfven crossing time of the inner magnetosphere, which takes $\sim 0.25 $ msec
\citep[][]{palmer05}. This unambiguously points to the
magnetospheric origin of GFs, presumably during reconnection event in the magnetically-dominated \ms\  \citep{lyutikov06}. 

The processes that cause magnetar X-ray flares (and possibly the persistent
emission) may be similar to those operating in the  solar corona. The energy of
magnetar flares is accumulated inside the neutron star at the moment of its
formation in the form of interior electric currents.  These currents are then
slowly pushed into the magnetosphere, gated by {slow, plastic deformations} of
the neutron star crust.  This leads to gradual twisting of the magnetospheric
field lines, on time scales much longer than  the  magnetar's GF, and
creates active magnetospheric regions similar to  the Sun's spots.  Initially, when the electric current (and possibly the magnetic flux)
is pushed from the interior of the star into  the \ms,  the latter slowly adjusts to the changing boundary
conditions.  As more and more current is pushed into the magnetosphere,
it eventually reaches a point of dynamical instability. The loss of stability leads to a
rapid restructuring of magnetic configuration, on the \Alfven crossing time
scale, to the formation of narrow current sheets, and to the  onset of magnetic dissipation.
As a result, a large amount of magnetic energy is converted into the kinetic and  bulk
motion  and radiation.

Observationally, a number of data point to the reconnection origin of the magnetar flares.
 Other predictions of the
model, confirmed by observations are (i) the post-flare magnetosphere has a
simpler structure, as the pre-flare network of currents has been largely
dissipated; (ii) its spectrum is softer, since the hardness of the spectrum is
a measure of the current strength in the bulk of magnetosphere
\citep{tlk02,lf05}, with softer spectra corresponding to a smaller external
current; and (iii) the spin-down rate, which depends on the amount of electric
current flowing through the open field lines, decreases.  The observations of
two recent GFs, in SGR 1900+14 and SGR 1806-20, fully agree with these
predictions.
In both cases the persistent flux increased by a factor of two,
its spectrum hardened (the power law index decreased from 2.2 to 1.5), and the
spin-down rate increased in the months leading to the flare \citep{turolla}.
In the post-flare period, the pulsed fraction and the spin-down rate have
significantly decreased and the spectrum softened \citep{rea05}. All these
effects imply an increase of the external current before, and a decrease
after  the flare \citep{lyutikov06}.

\subsection{Reconnection in magnetized jets of Active Galactic Nuclei and Gamma Ray Bursts}
 Recent observations of  AGNs in GeV and  TeV energy range have  raised new questions regarding the parameters of the central 
engine, and the location and kinematics of the high energy $\gamma$-ray, as well as X-ray and radio emission zones.  In particular, the rapid  flares
reported for Mrk 501 and PKS 2155-304, on timescales of 3-5 minutes \citep{2007ApJ...669..862A, 2007ApJ...664L..71A} imply  an emitting size smaller than the gravitational radius $t_{lc}\sim$hours of the supermassive black
holes of these blazars. 
This  indicates a very high Doppler factors $\delta$, exceeding $\delta =100$.  A similar  estimate also comes from the requirement that the TeV photons
escape the production region. 

While highly relativistic motion may appear to be a cure-all, the  bulk Lorentz factor $\Gamma$ can be directly constrained by VLBI observations of bright blobs moving with apparent speeds on the sky, $\beta_{app}$, that appear to be superluminal.  This type of motion occurs when the emitting region is moving relativistically and close to the line of sight \citep{Rees:1966}.  The apparent motion can exceed $c$ due to propagation effects.  If a blob is moving along with the bulk flow of a jet and its velocity vector makes an angle, $\theta_{ob}$, with the line of sight, then its apparent motion transverse to the line of sight will be:
$
\beta_{app}=\frac{\beta_{\Gamma}\sin{\theta_{ob}}}{1-\beta_{\Gamma}\cos{\theta_{ob}}}
$.
The maximum $\beta_{app}$ can reach is $\beta_{\Gamma} \Gamma$ when $\theta_{ob}\cong1/\Gamma$.  Thus, if the blob motion corresponds to the underlying bulk motion of the jet, measuring $\beta_{app}$ can constrain the possible bulk Lorentz factor, $\Gamma$.

Following the suggestion by  \cite{LyutikovRem},
a number of authors \citep{2009MNRAS.395L..29G,2008MNRAS.386L..28G,PiranTurb,2009MNRAS.395..472K} proposed that fast time scale variability both in AGNs and GRBs is produced by ''mini-jets'',  compact emitting 
regions that move relativistically {\it within} a jet of bulk $\Gamma \sim 10$ (in case of AGNs; for GRBs $\Gamma \sim 100$). Thus, the emission is beamed in the bulk outflow frame,  \eg\  due to relativistic motion of (using pulsar physics parlance)  ''fundamental emitters''. 

 In GRBs, claims of high polarization \citep{coburn03,Willis} offer direct measurements of the possibly  dominant large-scale \Bf. If confirmed, these observations argue in favor of magnetic reconnection as the main particle acceleration mechanism. 
 
\subsection{ Reconnection in the Double Pulsar system PSR J0737$-$3039}

Detection by \cite{mclau04} of drifting sub-pulses of pulsar B   in the Double Pulsar system PSR J0737$-$3039 with the frequency related to A period, presents an excellent opportunity to use Pulsar B as a probe of Pulsar A wind properties
at $\sim 1000$
light cylinder radii of A, many orders of magnitude closer that have been possible so far.   
In particular, the fact that the observed
modulation is at the frequency of A, and not a double  frequency, 
already can be used as an indication that a large fraction of A wind is carried by
relativistic MHD waves so that {\it directions} of electric and/or \Bfs are important, not only 
total pressure of the wind.

A possible explanation is that the modulation of B by A is due to reconnection between
\Bfs\ in the wind and in the B \ms.  When magnetospheric field lines connect to the wind's \Bf, they are ``dragged'' by the wind. Half a spin period of pulsar A later, when the wind's \Bf\ changes polarity, the magnetospheric \Bf\ disconnects from the wind and relaxes back to the position given by the   impenetrable conductive boundary conditions.  As the radio emission is produced along the local direction of a  \Bf, this periodic "dragging" and relaxation.

\section{Dynamic force-free plasma} 
In astrophysical settings 
the  local microscopic  plasma time scale (\eg plasma frequency)
 is much shorter
than the global  dynamical time scales
and 
there
is plenty of charges available to screen the component of
electric field along the magnetic field.
In addition, astrophysical plasma is usually  collisionless, making it 
an  extremely  good conductor. 
In addition, there
is plenty of charges available to screen the component of
electric field along the magnetic field  $\E \cdot \B=0$.
Even in the extreme cases, when plasma may not be able 
to short out parallel electric field due to lack of available changes 
(charge separated flows)
 various
radiative process (\eg  emission of curvature photon
or through inverse Compton scattering)
 may lead to what is known as vacuum  breakdown: abundant 
production of electron-positron pairs.  The newly born
pairs will create a charge density that would shut-off the accelerating
electric field.
The typical potential difference
$\Delta V_{{\rm vac}}\sim kT/e$ needed to break down the vacuum in a GRB
is typically in the MV-GV range. This is often orders of magnitude smaller than
typically available  EMF ($\sim 10^{15}-10^{16}$ eV for pulsars, $\sim 10^{18}-10^{20}$ eV for AGNs and  GRBs).

The properties of plasma in the magnetospheres of pulsars and magnetars, pulsar
winds, AGN and GRB jets are very different from those of more conventional Solar
and laboratory plasmas.  The principal difference is that it is relativistically
strongly magnetized.  In order to describe the level of magnetization it is
convenient to use the so-called magnetization parameter $\sigma = 2 ( {u_B /u_{p}})$,
where $u_B=B^2/8\pi$ is the magnetic energy density and $u_p= \rho c^2$ is the
rest mass-energy density. In traditional plasmas this parameter is very small.
On the contrary,  in  some astrophysical settings it is likely to be very large.
For example, in magnetars \be {\om_B R_{NS} \over c} \left({m_e \over m_p}
\right) \sim 10^{13} \leq \sigma \leq {\om_B \over \Omega} \left({m_e \over m_p}
\right) \sim 10^{16} \ee (the upper limit corresponds to the Goldreich-Julian
density of electron-ion plasma whereas the lower limit corresponds to the
poloidal current producing the toroidal \Bfs of the same order as the poloidal
one.  Here $\om_B = e B/m_e c$ is the cyclotron frequency, $B$ is the \Bf at the
\NS\ surface, $R_{NS} $ is the \NS\ radius, and $\Omega$ is its rotational
frequency.

The parameter regime of highly magnetized plasma, $\sigma \gg 1$, implies that (i) the inertia of this
plasma is dominated by the \Bf and not by the particle rest mass, $B^2/ 8 \pi
\gg \rho c^2$, (ii) the propagation speed of \Alfven waves approaches the speed
of light, (iii) the conduction current flows mostly 
along the \Bf lines, (iv) the
displacement current $(c / 4 \pi) \partial_t {\bf E}$ may be of the same order
as the conduction current, ${\bf j}$, (v) the electric charge density, $\rho_e$,
may be of the order of $j/c$.  These are very different from the properties of
laboratory plasmas, plasmas of planetary magnetospheres, and the interplanetary
plasma, the cases where  plenty of experimental data and theoretical
results exist.

The large expected value of $\sigma$ (or small $1/\sigma$) may be used as an
expansion parameter in the equations of relativistic magnetohydrodynamics. The
zero order equations describe the so-called relativistic force-free
approximation.  One may see this limit as the model where massless charged
particles support currents and charge densities such that the total Lorentz
force vanishes all the time (this also insures the ideal condition ${\bf E}
\cdot {\bf B}=0$.)  
This  allows one to related  the current to electro-magnetic fields \citep{Gruzinov99}
\be
{\bf J}= {c \over 4 \pi} {({\bf E}\times{\bf B})\nabla\cdot{\bf E}+
({\bf B}\cdot\nabla\times{\bf B}-{\bf E}\cdot
\nabla\times{\bf E}){\bf B}\over   B^2}
\label{FF}
\ee
This may be considered as the Ohm's law for relativistic force-free electro-dynamics.
(Note that this implies that the invariant
${\bf E}\cdot{\bf B}=0$ and that electromagnetic energy
is conserved, ${\bf E}\cdot{\bf J}=0$.)

Under stationary and axisymmetric conditions, these equations guarantee
that the angular velocity $\Omega$ is conserved along field lines.
They also require a space charge density $\rho=\nabla\cdot{\bf E}/(4 \pi)$
of magnitude $\sim\Omega B/c$ to develop. (Formally this, like the equation
$\nabla\cdot{\bf B}=0$, is just an initial condition.)

In the non-relativistic  plasma the notion of force-free fields is often related
to the stationary configuration attained asymptotically  by the system
(subject to some boundary conditions and some constraints, \eg conservation 
of helicity). This equilibrium is attained on time scales of the order of the
\Alfven crossing times. In  strongly magnetized 
relativistic plasma the \Alfven speed may become of the order of the 
speed of light $c$, so that 
 crossing times becomes of the order of the light travel time. But if plasma
is moving relativistically its state is changing on the same time scale. 
This leads to a notion of dynamical force-free fields.

The force-free condition can be re-expressed by setting the divergence of the
electromagnetic stress tensor to zero. This form has the merit that
it brings out the analogy with fluid mechanics. Electromagnetic stress
pushes and pulls electromagnetic energy which moves with
an electromagnetic velocity ${\bf E}\times{\bf B}/B^2$, perpendicular
to the electric and magnetic fields. This is the velocity of the
frames (only defined up to an arbitrary Lorenz boost along
the magnetic field direction)
in which the electric field vanishes, (provided that the first electromagnetic invariant
$B^2-E^2>0$).

The limit $\sigma \rightarrow \infty$ is somewhat
reminiscent of  subsonic hydrodynamics as both the fast speed and the \Alfven
speed approach the speed of light. 
For example, in case of slow processes, taking place on time scales much longer
than light travel time, the Maxwell's equations may be written as continuity and momentum conservation \citep{2007MNRAS.374..415K}):
\begin{equation}
\partial_t{\rho}+\nabla(2\rho{\bf V})=0, \hskip .5 truein \partial_t{\rho{\bf
V}} + \nabla\left( -\frac{{\bf B}\otimes{\bf B}}{4\pi} + {\bf I}
\frac{B^2}{8\pi}\right) =0.
\label{cont}
\end{equation}
where $\rho={B^2}/{8\pi c^2}$ is the effective mass density of the
electromagnetic field and ${\bf V} = {\bf E} \times {\bf B} /B^2$ is the drift
speed of charged particles, This closed system of equations is very similar to
non-relativistic MHD. This observation provides an interesting insight into the
dynamics of two very different dynamical systems.

\subsection{Time-dependent  hyperbolic  Grad-Shafranov equations}

 \cite{2011PhRvD..83l4035L}  demonstrated that the
 time evolution of the axisymmetric force-free magnetic  fields can be expressed in terms of  the 
 hyperbolic   Grad-Shafranov equation,  under the  assumption that the fields remain axially-symmetric. Qualitatively, there two separate types of non-stationarity: (i) due to the variations of the current $I(t)$ for a given shape of the flux function; (ii)  due to the variations of the shape of the flux function  for a given current $I$.
Using these equations it is possible to   find exact non-linear  time-dependent  
 Michel-type  (split-monopole) structure of \mss, \eg,  driven by spinning and collapsing \NS\ in \Sc\ geometry:
  \ba &&
  B_r = \left( {R_{s} \over r} \right)^2 B_s, \, B_\phi = - { R_{s}^2 \Omega \sin \theta \over r} B_s, \,  E_\theta = B_\phi
  \nn &&
j_r =  -2  \left( {R_{s} \over r} \right)^2  \cos  \theta \Omega B_s 
\nn && 
P= (1-\cos\theta) B_s  R_{s}^2
\nn &&
\Phi= - P \Omega
\nn  &&
I =-{  P (P-2 B_s  R_{s}^2)  \Omega \over 2 B_s  R_{s}^2} = {1\over 2} B_s  R_{s}^2 \Omega \sin ^2\theta
\label{FFF}
\ea
where $P$ is the flux function, and $\Phi$ is the electric potential and $\Omega = \Omega(r-t)$ is {\it an arbitrary function}.
Thus, we found exact solutions for time-dependent non-linear relativistic force-free configurations.
Though the configuration  is non-stationary (there is a time-dependent propagating
wave),  the form
of the flux surfaces remains constant.

\subsection{Limitations of force-free approach}

The generic limitation of the force-free formulation of MHD is that  the evolution of the
electromagnetic
field leads, under certain conditions,  to the formation of regions with $E>B$ \citep[\eg][]{2003ApJ...598..446U}, since there is no
mathematical limitation on $B^2-E^2$ changing a  sign under  a strict
force-free conditions.
 In practice, the particles in these regions  are subject to rapid acceleration
through $\vec E\times\vec B$ drift, following by a formation of pair plasma via various radiative effects and reduction of the \Ef. Thus, regions with $E>B$ are necessarily resistive. This  breaks the ideal assumption and leads to the  slippage of \Bf\ lines with respect to plasma.
In addition, evolution of the  magnetized plasma often  leads to formation of resistive current sheets, with the similar effect on \Bf.

As a simple example demonstrating that  the dynamical system described by the relativistic  force-free limit has a natural tendency to violate the physical requirement of a negative  first electromagnetic invariant, $E^2 - B^2  <0$, consider  a resistive  decay of a line current. Suppose at time $t=0-$ there is a line current $I_0$ that decays for  times $t>0$ according to $I =I_0 (1-t/\tau)$. This launches an outgoing rarefaction wave in which the EM field are given by 
 \ba &&
 B_\phi = {I_0 \over 2  \pi r} \left(1 - {\sqrt{t^2 -r^2} \over \tau } \right)
 \nn &&
 E_z = -  {I_0 \over 2  \pi  \tau} \ln { t- \sqrt{t^2 -r^2} \over r}
 \ea
 (for $r< t$). 
 The resulting radial inward velocity, $v_r = E_z/B_\phi$ at some moment becomes larger than the speed of light. This example illustrates an important point:  in force-free approximation plasma tends to develop dissipative regions where $E \rightarrow B$. This regions can develop non-locally, far from the cause that initiated the plasma motion (resistive decay at $r=0$). 
 
Another generic limitation of the force-free approach is related to the structure of the current sheet. Since in  the force-free limit the plasma pressures are neglected, nothing prevents formation of very thing  and thus highly dissipative current sheets. As a result, numerical models then can produce exceptionally high dissipation rates \citep[\eg][]{2012arXiv1209.5121G,2011arXiv1112.2622L}. If small kinetic pressure in taken into account \cite[\eg][]{2011PhRvD..84h4019L,2012arXiv1211.2803T}, the reconnection rate drops considerably. 
Another approach that allows (partial) stabilization of the current sheet is the resistive force-free plasma dynamics, which we discuss next.

\section{ Dissipation in highly magnetized plasmas }
\label{Dissipation}

\subsection{ Tearing mode in force-free plasma}
\label{DDDD}

One of the most important resistive instabilities in a conventional plasma is
the so-called tearing instability. This is one of the principle unstable
resistive modes, which plays a key role in various TOKAMAK discharges like the
sawtooth oscillations and the major disruptions \citep[\eg][]{Kadomtsev75}, and
leads to the unsteady reconnection of Solar flares
\citep[\eg][]{Shivamoggi85,Aschwanden02} and Earth's magnetotail
\citep[\eg][]{gca78}.  The most important property of the tearing instability is
the growth time that is much shorter than the resistive time.

In addition to being magnetically-dominated,
 microscopic plasma processes, like particle collisions or 
 plasma turbulence, may
   contribute to resistivity and thus  make plasma  non-ideal.
 Resistivity will result in the decay of currents supporting the magnetic field; this, in turn,
will influence the plasma dynamics. 
Introduction of resistivity into force-free formulation is not entirely self-consistent. The reason is 
that in force-free plasma the velocity along the field is not
 defined. Since  plasma resistivity must be defined in the plasma rest-frame
this creates a principal ambiguity.

The force-free tearing mode has been considered by \cite{l03} \citep[see also][for a somewhat different formulation of resistive force-free plasma]{2007arXiv0710.1875G,2012ApJ...746...60L}.  \cite{l03} found that
similar to the non-relativistic case, the resistive force-free current layers
are unstable toward the formation of small-scale dissipative current sheets.  He
has also found that the growth rate of tearing instability, 
$\tau=\sqrt{\tau_d\tau_a}$, is intermediate
between the short \Alfven time scale $ \tau_a$ (which equals to 
the light crossing time in the force-free regime) and the long 
resistive time scale $\tau_d=l^2/\eta$, where $l$ is the width of the 
current layer. This is exactly the same expression as in the
non-relativistic case, which is rather surprising given the fact that 
the dynamic equations of force-free plasma are very different from 
the equations of non-relativistic MHD. 

\begin{figure}[h]
 \includegraphics[width=0.3\linewidth]{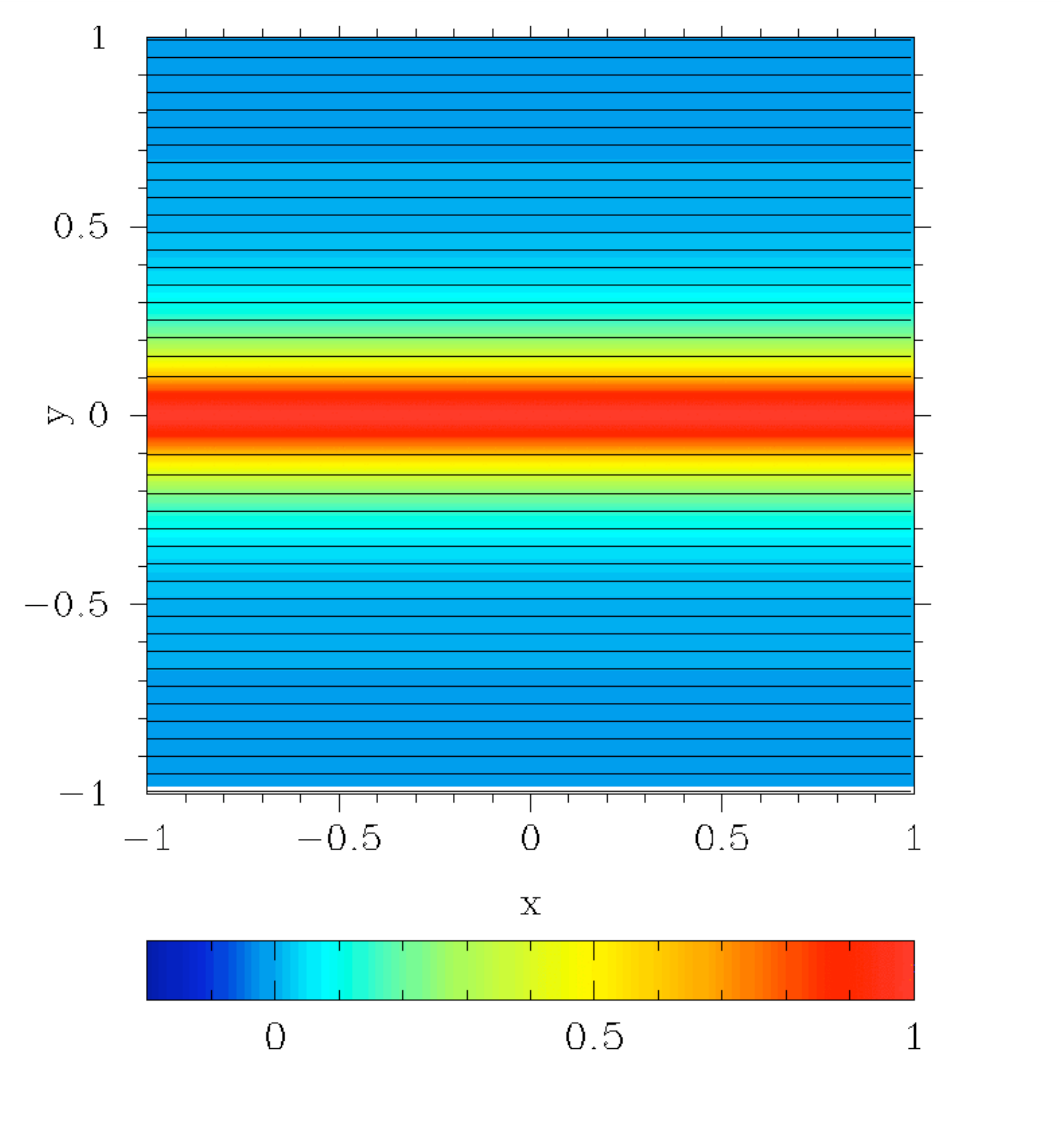}
\includegraphics[width=0.3\linewidth]{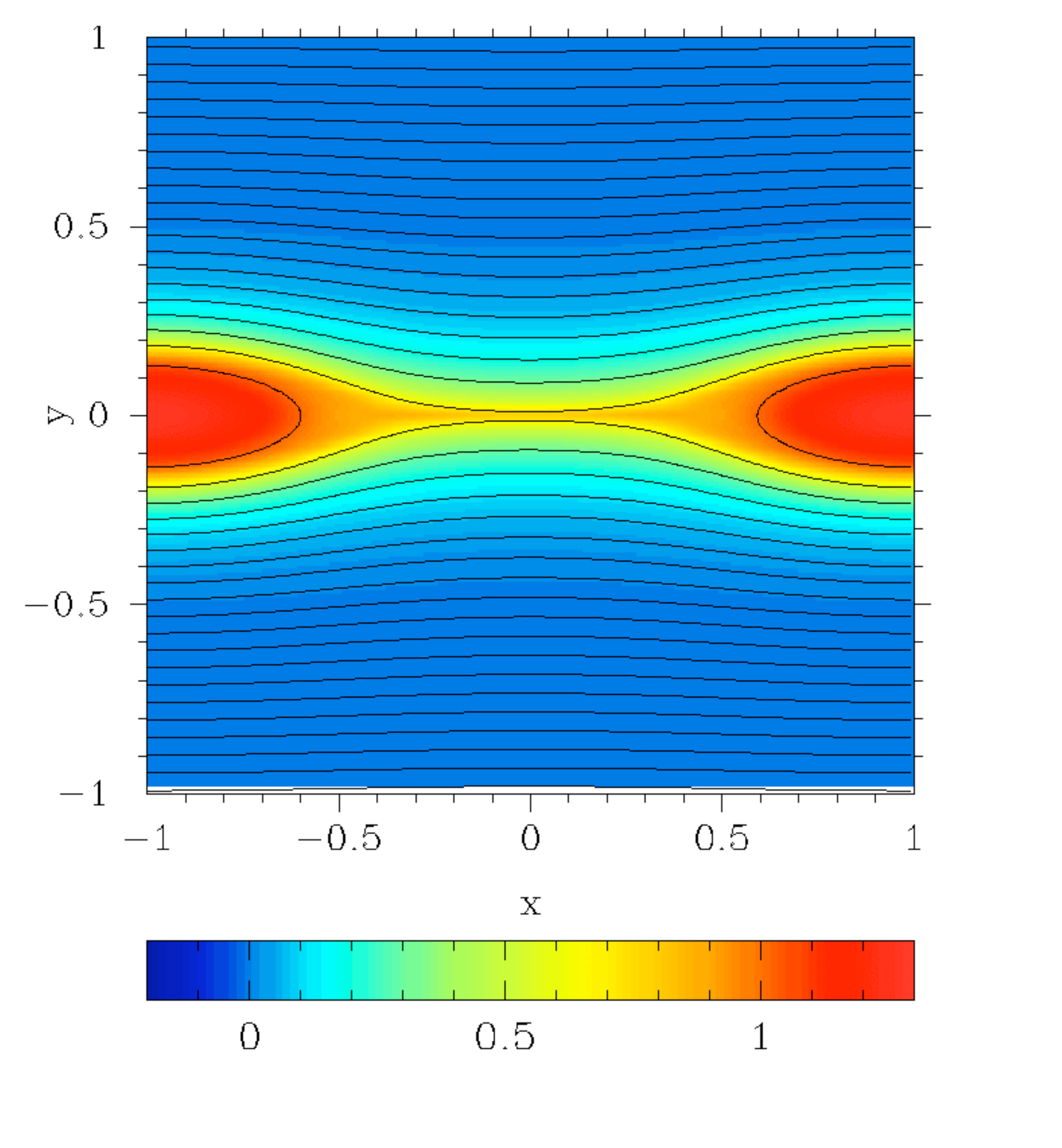}
  \includegraphics[width=0.3\linewidth]{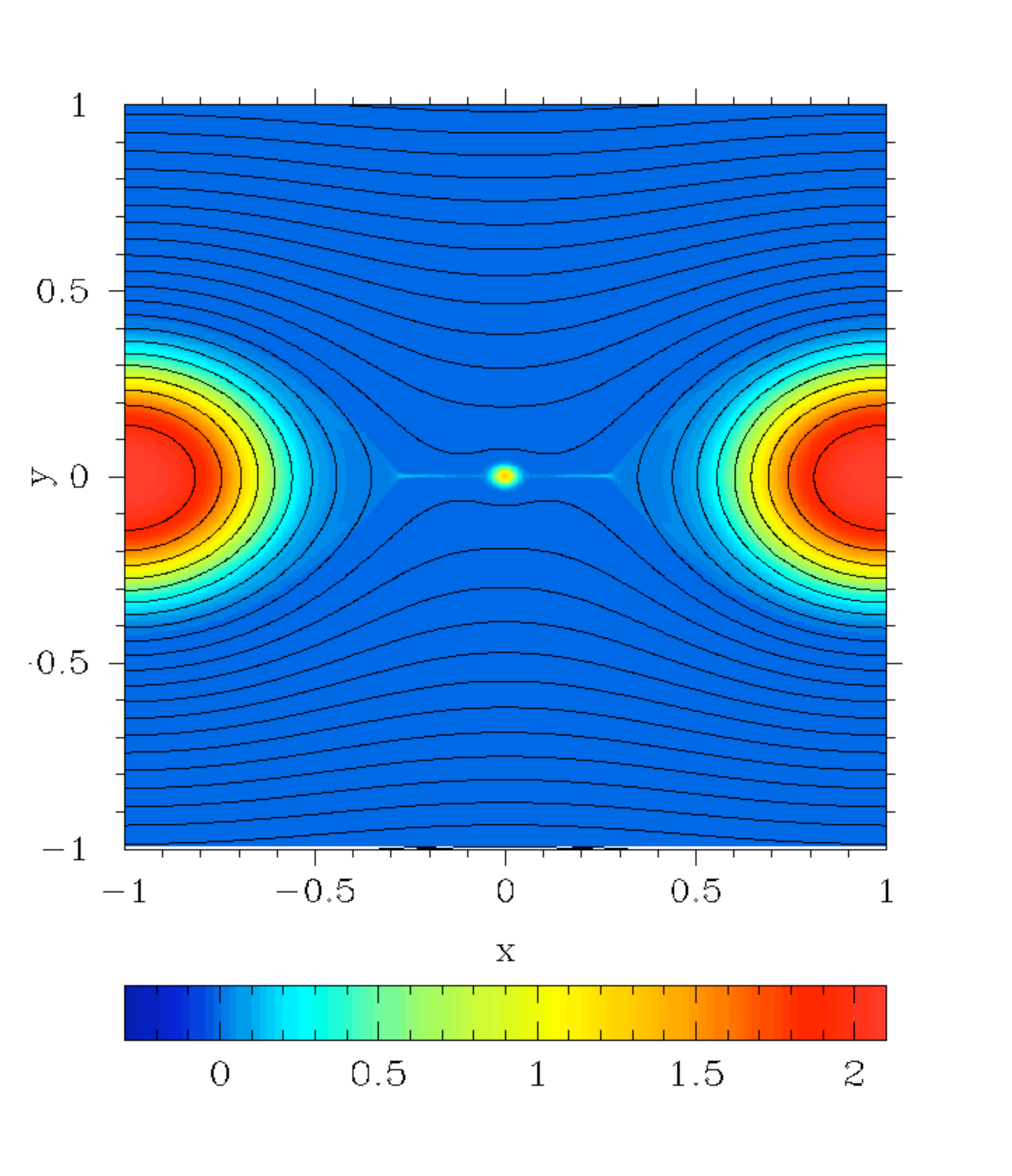}
\caption{Numerical simulations of the tearing
instability in force-free plasma. The color images show the distribution of
$B_z$ and the contours show the magnetic field lines. In this simulations the
perturbation has the wavelength corresponding to the maximum growth rate, $\tau
\simeq \sqrt{\tau_d\tau_c}$, where $\tau_d$ is the resistive time and $\tau_c$
is the light crossing time of the current layer. The Lundquist number
$L_u=\tau_d/\tau_c=10^3$.}
\label{tearing}
\end{figure}

Numerical modeling of the tearing instability in strongly magnetized plasma \cite{2007MNRAS.374..415K} (see Fig. \ref{tearing}.) fully  confirm the analytical estimates.

In high Reynolds number plasma it is expected that tearing developing into turbulence. This has been observed in both PIC and MHD recent simulations (see below).

\subsection{X-point collapse in force-free plasma}
\label{X-point}

Consider a vicinity of an X-point. The non-current-carrying configuration  has null lines intersecting at 90 degrees.
Following the work on the collapse of a non-relativistic X-point \citep{1953MNRAS.113..180D,1967JETP...25..656I,priest00}, let us assume that the initial configuration is squeezed by a factor $\lambda$, so that the initial configuration has a vector potential $A_z \propto x^2 - y^2/\lambda^2$. In addition, we assume that there is an axial constant \Bf\ $B_z$. 

We  are looking for time evolution of a vector potential of the type 
\be
A_z =-  \left( {x^2 \over a(t)^2} -{y^2\over b(t)^2} \right) { B_0 \over 2 L},
\ee
where parameters $B_0$ and $L$ charachterize the overall scaling of the \Bf\ and the spacial scale of the problem.

The initial condition for the squeezed X-point and the ideal condition then require
\ba &&
b(t) = \lambda /a(t)
\nn &&
\Phi = x y { B_z\over c}  \partial_t \ln a
\ea
(Since $\Delta \Phi=0$ there is no  induced charge density.) Parameter $a$ characterizes the ``squeezeness'' of the configuration; $a=1$ is the current-free case.

Faraday's law is then an identity, while the induction equation  in the limit $x,y \rightarrow 0$ gives
\be
\partial_t^2    \ln a  ={\cal A} \left( { a^4-\lambda^2 \over \lambda^4} \right), \,
 {\cal A}  =  { c^2 \over L^2} {B_0^2 \over B_z^2}
\label{addot}
 \ee

\begin{figure}[h!]
\centering
\includegraphics[width=.45\textwidth]{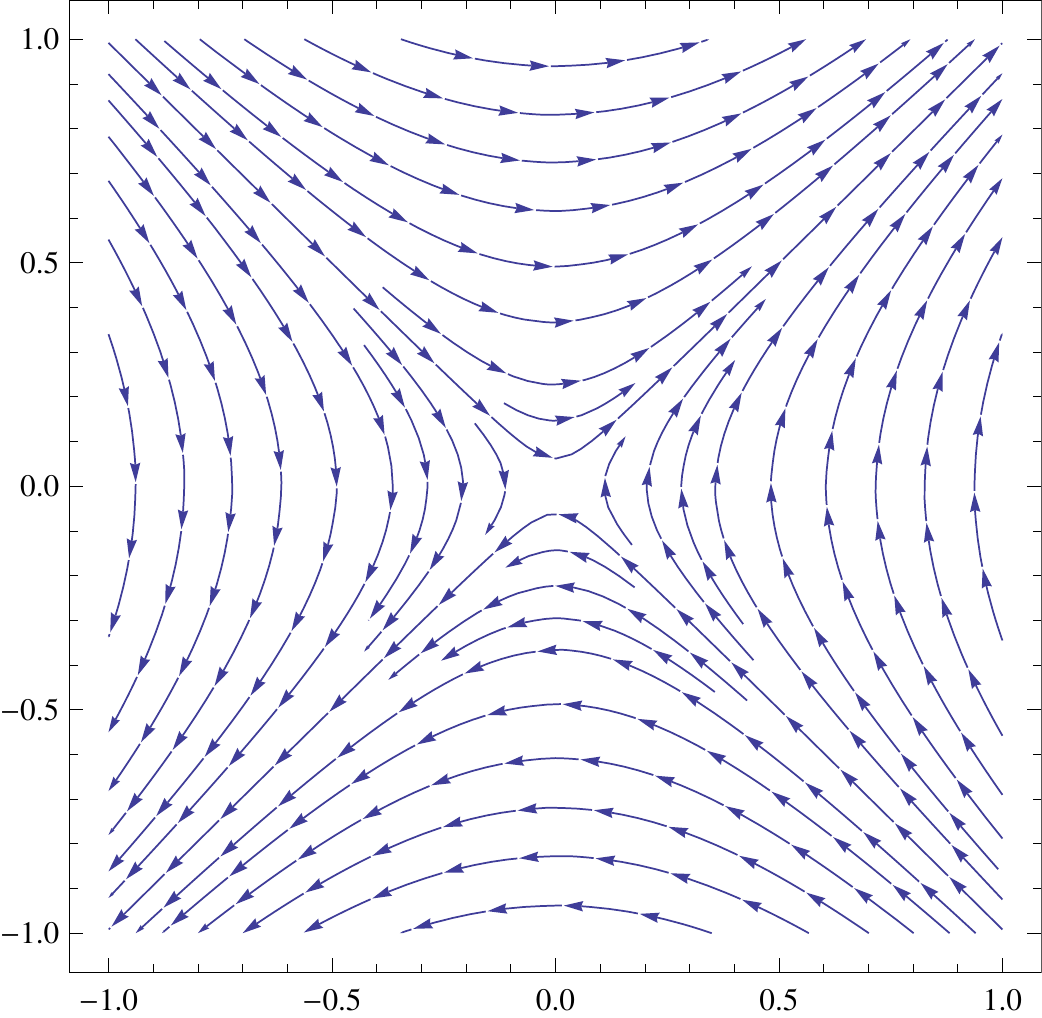}
\includegraphics[width=.45\textwidth]{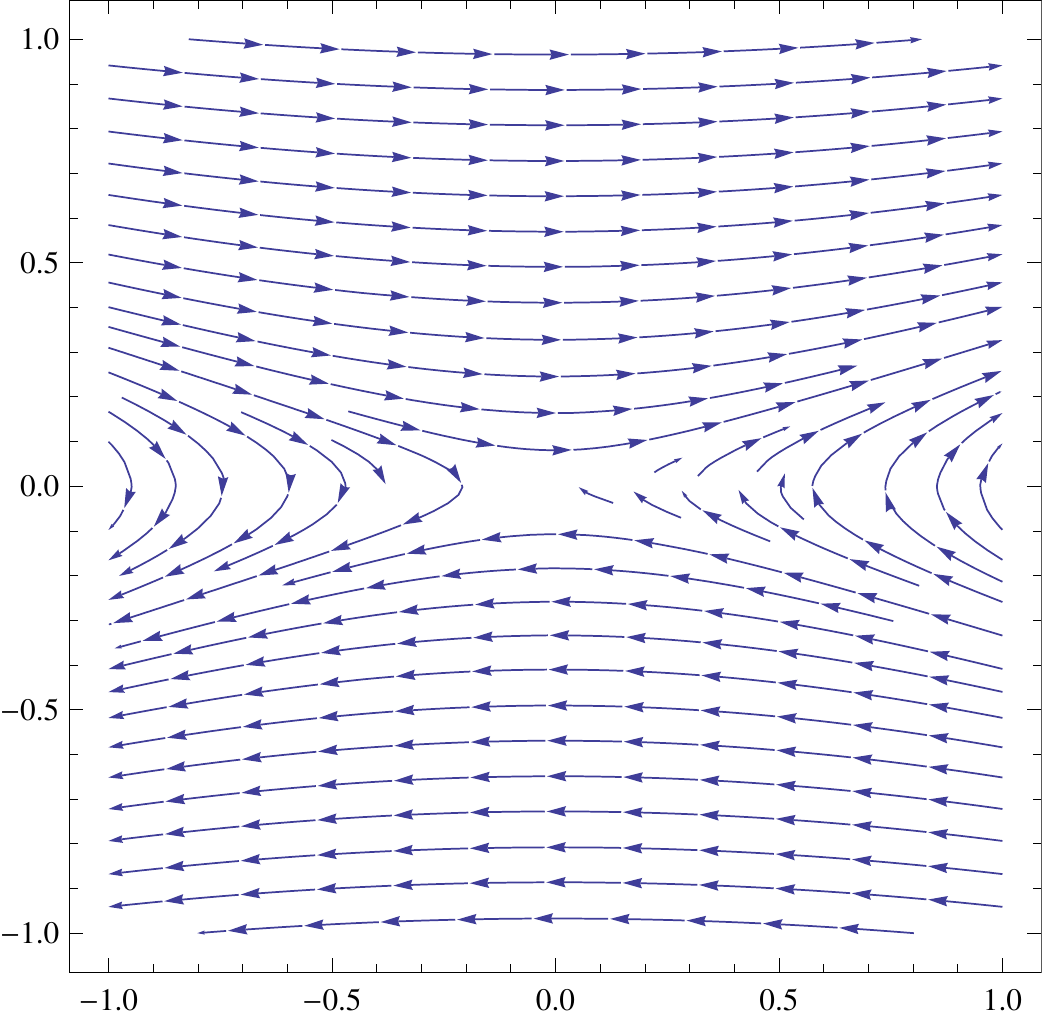}
\caption{Structure of the \Bf\ in the $x-y$ plane during X-point collapse in force-free plasma. The initial configuration on the left is slightly ``squeezed''. On dynamical time scale the X-point collapses to form  a current sheet, right figure. The structure of the \Ef\ in the $x-y$ plane  does not change during the collapse and qualitatively  resembles the $t=0$ configuration of the \Bf.}
\label{B-field-coll} 
\end{figure}

 Solutions of the equations  (\ref{addot}) show that $a(t)$ has  a finite time singularity for $\lambda < 1$: in finite time $a$ becomes infinite. At the moment  when one of the parameters $a$ or $b$ becomes zero, the  current sheet forms,  see Fig. \ref{B-field-coll}. 

  For small times $t \rightarrow 0$ (when $|a-1| << 1$),  with initial conditions $a(0)=1, \, a'(0)=0$, and assuming that the initial ``squeezing'' is small, $ \lambda=1-\epsilon,\, \epsilon \ll 1$, the solution is
 \be
a=1 + \epsilon \sinh ^2 \left( {  {c t \over L} {B_0 \over B_z} }\right) 
 \ee
 Thus,  the typical collapse time is 
 \be
 \tau \sim {L \over c} {B_z \over B_0}
 \ee
 is of the order of the \Alfven (light) crossing time of the initial configuration. 
 At these early times the \Ef\ grows exponentially
  
 At early times the particle drift follows a trajectory in the $x-y$ plane $ y \propto 1/x$. During the final collapse, in the limit $a \rightarrow \infty$, the particle distribution is further squeezed towards the neutral layer, $y\propto 1/(a^2 \sqrt{\ln x})$ (though in this limit the drift approximation becomes inapplicable.)
This shows that the X-point is stretched in one direction and is compressed in the other direction (note that $\nabla \v=0$ -  collapse is incompressible at the initial stage).

Thus, in an ideal relativistic force-free plasma the X-point undergoes a finite time collapse. At the same time, particles are squeezed by the \EM\ drift towards the neutral layer.
 The assumption of the force-free plasma will be broken down when the inflow velocity would become of the order of the \Alfven velocity. Then, the maximum \Ef\ is $E \sim \beta_A B_0$ which is of the order of $B_0$, \Bf\ in the bulk,  for $\sigma \geq 1$. 
The collapse occurs faster for small axial field $B_z \leq B_0$.

\subsection{Stationary  relativistic reconnection}

Magnetic reconnection is widely recognized as a very important phenomenon in
many laboratory and astrophysical plasmas (Biskamp 2000, Priest \& Forbes
2000). It has been studied very extensively over the last 40 years, and a 
significant progress has been made. However,
historically, the main applications of the reconnection theory were confined 
to Solar physics, Earth's magnetosphere and the fusion projects. 
In all these cases, the magnetic energy density is much smaller than the particle 
rest mass-energy density, and the characteristic speeds are much less than  
the speed of light. Therefore it is not surprising that most of the progress
has been made in the non-relativistic regime.

The analytical studies of dissipative effects in resistive magnetically dominated
plasmas, though limited in their generality,  provide an important first step
towards the full understanding of the plasma dynamics under these extreme conditions 
and will serve as a guide for numerical investigations. 

The first step is the generalization of non-relativistic models to the new regime \cite{LyutikovUzdensky,2005MNRAS.358..113L}
According to \cite{LyutikovUzdensky}, the relativistic theory of
the simplest model of magnetic reconnection --- the Sweet--Parker model --- 
involves two very large parameters: 
the Lundquist number, $L_u$, and the magnetization 
parameter $\sigma$. The simplest  Sweet--Parker model of relativistic reconnection cannot be built self-consistently. The reason is that the convention non-relativistic model operates only with conservation laws, and not with the dynamical structure. In the relativistic case the conservation of energy then predicts acceleration to high Lorentz factors \citep{LyutikovUzdensky}, but  this would lead to large pressure drop and would violate a force-balance across the current sheet \citep{2005MNRAS.358..113L}. 

Overall properties of stationary relativistic reconnection  (like inflow and outflow velocities, Sweet-Parker versus Petschek models) remains an open question 
\citep[for a recently review see][]{2012SSRv..tmp...78H}.
The shift in our understanding of non-relativistic reconnection layers, in particular as we discuss in more detail later, the importance of turbulence (Lazarian \& Vishniac 1999) that is present in 
most astrophysical systems due to numerous instabilities that prey on high
Reynolds number velocity fluids. 

Interestingly enough, reconnection itself
creates turbulence, which can be the cause of ``reconnection instability'' described
in Lazarian \& Vishniac (1999, 2009), which develops when the initial level of turbulence in the system is low or even the magnetic fields are originally laminar. As the outflow gets turbulent, the level of turbulence
in the system and the reconnection rate increases inducing the positive feedback. This process may result in bursty reconnection of the time seen in solar flares, where the initial state of turbulence is low. This process can be also a driver of other dramatic energy bursts, e.g. gamma ray bursts (Lazarian et al. 2003,
Zhang \& Yan 2011, Lazarian \& Yan 2012).

It is important that the initially laminar reconnection layer is subject to  
 the tearing mode instability\footnote{The estimates of the tearing mode in highly relativistic plasmas (see \S \ref{DDDD}) indicate that the mode growth rate (surprisingly)  follows the non-relativistic scaling.} \cite{2007PhPl...14j0703L}, which both drive reconnection and ensure the turbulent state of the 3D reconnection layer (see Karimabadi 2013, Lazarian \& Karimabadi 2013, Beresnyak 2013).
All this puts in doubt many stationary and laminar reconnection models. Indeed,  as described above, in astrophysics the relativistic reconnection is invoked for highly non-stationary processes and a laminar state of astrophysical fluid is more of an exception rather than a rule.  
Thus, one might expect that in the relativistic regime the current sheet is being fragmented and broadened by turbulence or/and subject to tearing.

\subsubsection{Reconnection in the presence of turbulence}

Properties of fluids are know to be strongly affected by turbulence. For instance, diffusion in turbulent fluids does
not depend on molecular diffusivity. Thus, it is important to understand what can be the role of turbulence for
the diffusion of magnetic field and reconnection that this transport can entail within relativistic plasmas. Below we provide 
arguments suggesting that turbulence can make magnetic reconnection fast. 

\begin{figure}[h!]
\centering
\includegraphics[width=.85\textwidth]{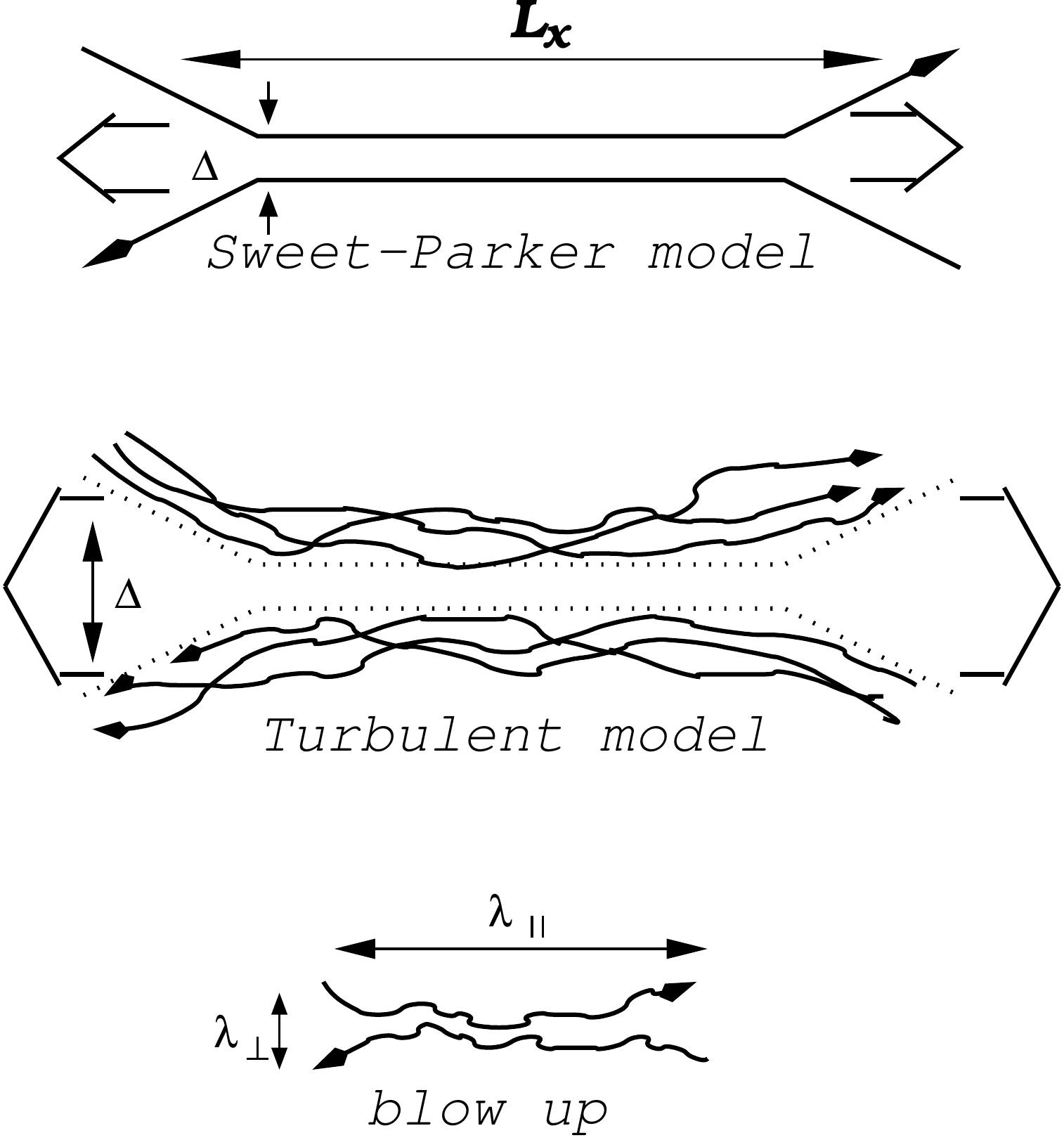}
\caption{Upper plot: Sweet-Parker model of reconnection. The outflow is limited by a thin slot Æ, which is determined by Ohmic diffusivity. The other scale is an astrophysical scale Lx>>Æ. Middle  plot: Reconnection of weakly stochastic magnetic field according to LV99. The Goldreich-Sridhar (1995) model of MHD turbulence is used to account for the stochasticity of magnetic field lines. The outflow in the LV99 theory is limited by the diffusion of magnetic field lines, which depends on field line stochasticity. Lower plot: An individual small-scale reconnection region. The reconnection over small patches of magnetic field determines the local reconnection rate. The global reconnection rate is substantially larger as many independent patches come together. From Lazarian et al. (2004).}
\label{figure1}
\end{figure}

In terms of non-relativistic fluids, the predictive model of turbulent reconnection was presented in Lazarian \& Vishniac (1999) [henceforth LV99]. 
LV99 considered reconnection in the presence of sub-{\Alfven}ic turbulence in magnetized plasmas. They identified stochastic wandering of magnetic field-lines as the most critical property of MHD turbulence which permits fast reconnection.  As illustrated in Figure\ref{figure1}, this line-wandering widens the outflow region and alleviates the controlling constraint of mass conservation\footnote{The LV99 model is radically different from its predecessors which also appealed to the effects of turbulence. For instance, unlike Speiser (1970) and Jacobson \& Moses (1984) the model does not appeal to changes of microscopic properties of plasma.  The nearest progenitor to LV99 was the work of Matthaeus \& Lamkin (1985) Matthaeus \& Lamkin (1986), who studied the problem numerically in 2D MHD and who suggested that magnetic reconnection may be fast due to a number of turbulence effects, e.g. multiple X points and turbulent EMF. However, these papers did not address the important role of magnetic field-line wandering,  and did not obtain a quantitative prediction for the reconnection rate, as did LV99.} 

One can argue that the LV99 model that was shown to make reconnection fast carries over to relativistic fluids the same way as other, e.g. Sweet-Parker model does.  In fact, it is clear from Figure \ref{figure1} that LV99 generalizes the Sweet-Parker model for the case of turbulent magnetic fields. The
limitation that makes the Sweet-Parker reconnection slow both in relativistic and non-relativistic cases stems from the fact that the thickness of the outflow region is limited by effects related to plasma conductivity. These effects are related to microscopic scales and make $\Delta\ll L$ for the case of laminar fluids. In the case of
LV99 reconnection the outflow is determined by macroscopic field wandering and therefore the $\Delta$ can get comparable with $L$. As the mass conservation
dictates that the reconnection velocity is 
\begin{equation}
V_{rec}\approx V_A  \Delta/L
\label{Vrec}
\end{equation}
LV99 reconnection gets fast and it depends only on the intensity and injection scale of turbulence. In the relativistic limit $V_A$ approaches the velocity of light and therefore one can expect Eq. (\ref{Vrec}) to hold with the change of $V_A$ to $c$. At the same time, the exact scaling of $\Delta$ depends on the properties of relativistic turbulence. In any case, $\Delta$ if it determined by turbulence does not depend on microscopic plasma resistivity and therefore the reconnection should be fast\footnote{Magnetic reconnection is universally defined as fast when it does not depend on resistivity.} even if the scaling of turbulence in
relativistic case differs from non-relativistic one. 

On the basis of the recent studies of relativistic turbulence one may argue that LV99 model may be even directly applicable to the relativistic case. Indeed, the existing studies of the scaling and anisotropies of MHD turbulence (Cho 2008 and Cho \& Lazarian 2012) testify that the relation between the parallel and perpendicular scales of eddies as well as the spectrum of turbulence are the same for relativistic and non-relativistic turbulence. As $\Delta$ in LV99 theory is determined by those properties of turbulence one can argue that the LV99 expressions for the reconnection rates can be directly relevant to relativistic reconnection\footnote{One may question the application of the field wondering concept to relativistic case. However, LV99 expression for $\Delta$ (and therefore also reconnection rates) was re-derived in ELV11 without appealing to this concept.}.
This can be used at least as an educated guess for the discussion that we present further. 

Irrespectively of the exact correspondence of the properties of relativistic and non-relativistic MHD turbulence one can argue that in the presence of turbulence
$\Delta$ should increase and therefore the rate of magnetic reconnection should grow. This provides a prediction of flares of reconnection, explaining bursty
energy release that is observed in solar flares as well as in many high energy phenomena, as it discussed in LV99. Indeed, if magnetic fluxes in contact are initially laminar or very weakly turbulent, $\Delta$ and therefore the reconnection rate may be slow initially. However this situation is unstable in the sense that if the outflow of plasma from the reconnection region gets turbulent, this will increase the turbulence of the ambient magnetic field and increase $\Delta$. With larger $\Delta$
the outflow has larger Reynolds number and thus will get more turbulent. This results in "reconnection instability" (see more in Lazarian \& Vishniac 2010). Similarly,
one can argue that another process predicted in LV99 and reported in the observations by Sych et al. (2010), i.e. the initiation of reconnection by magnetic 
reconnection in adjacent regions should also be present in the relativistic case.

\subsubsection{Acceleration at relativistic reconnection}

\begin{figure*}[!t]
\includegraphics[width=\textwidth]{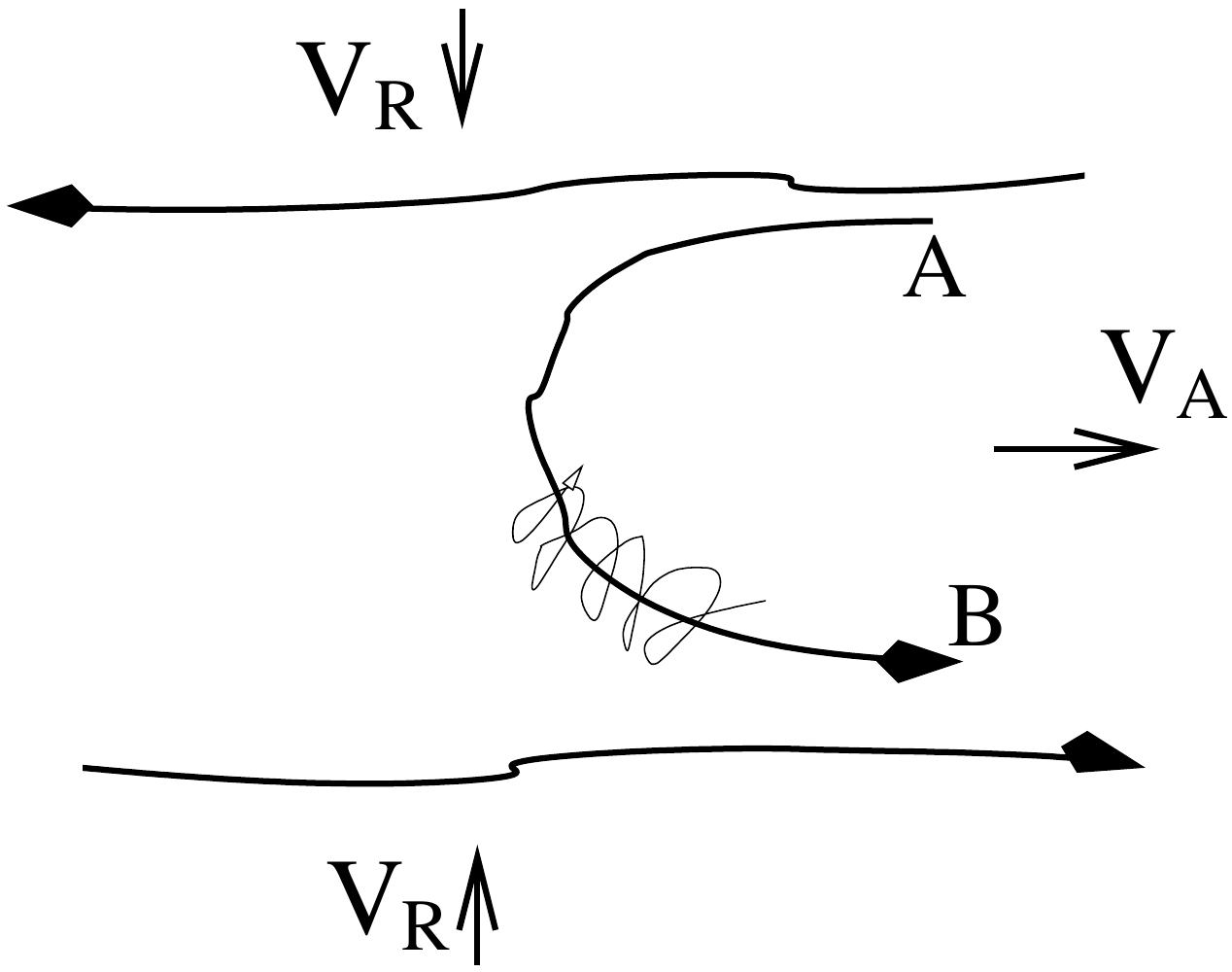}
\caption{Cosmic rays spiral about a reconnected magnetic field line and bounce
back at points A and B. The reconnected regions move towards each other with the
reconnection velocity $V_R$.   From Lazarian 2005.} 
\label{fig_accel1}
\end{figure*}

Studies of the First order Fermi acceleration has been performed so far for non-relativistic
reconnection (see de Gouveia dal Pino \& Lazarian 2005, Lazarian 2005, Drake et al. 2006,
Lazarian \& Opher 2009, Drake et al. 2010, Lazarian \& Desiati 2010, Lazarian et al. 2011, Kowal et al. 2012).  We briefly summarize those below.
 The first order acceleration of particles entrained on the contracting magnetic
loop can be understood from the Liouville theorem. In the process of the
magnetic tubes contraction a regular increase of the particle's energies is
expected. The requirement for the process to proceed efficiently is to keep the
accelerated particles within the contracting magnetic loop. This introduces
limitations on the particle diffusivity perpendicular to the magnetic field
direction. The subtlety of the point above is related to the fact that while in
the first-order Fermi acceleration in shocks magnetic compression is important,
the acceleration via the LV99 reconnection process is applicable even to
incompressible fluids. Thus, unlike shocks, it is not the entire volume that
shrinks for the acceleration, but only the volume of the magnetic flux tube.
Thus high perpendicular diffusion of particles may decouple them from the
magnetic field. Indeed, it is easy to see that while the particles within a
magnetic flux rope depicted in Figure~\ref{fig_accel1} bounce back and forth
between the converging mirrors and get accelerated, if these particles leave the
flux rope fast, they may start bouncing between the magnetic fields of different
flux ropes which may sometimes decrease their energy. Thus it is important that
the particle diffusion both in the parallel and perpendicular directions to the
magnetic field stay different. The particle anisotropy which arises from
particles preferentially getting acceleration in terms of the parallel momentum
may also be important. 

\begin{figure}[!t]
\includegraphics[width=\columnwidth]{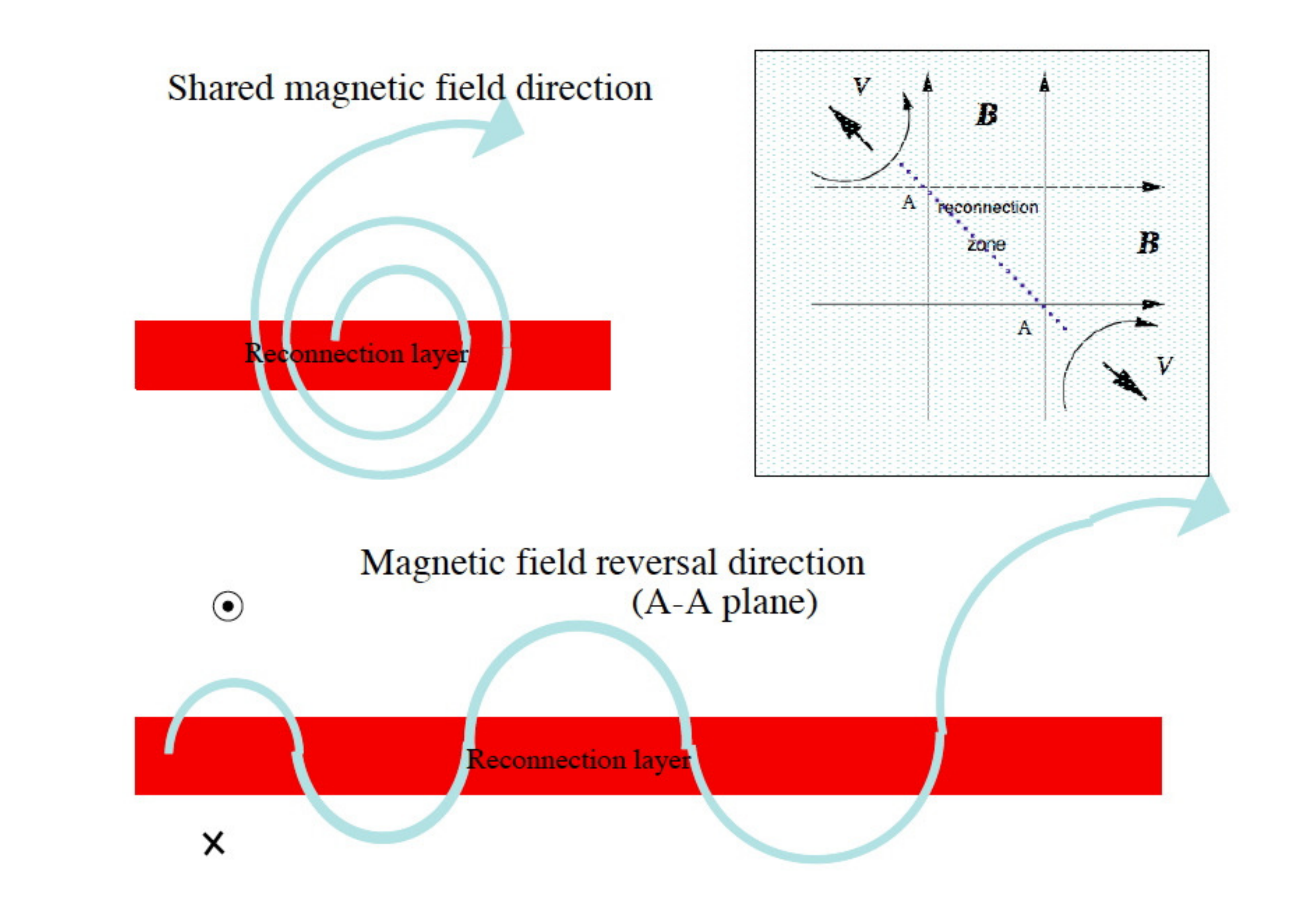}
\caption{Particles with a large Larmor radius gyrate about the magnetic field
shared by two reconnecting fluxes (the latter is frequently referred to as
``guide field''. As the particle interacts with converging magnetized flow
corresponding to the reconnecting components of magnetic field, the particle
gets energy gain during every gyration. From Lazarian et al. 2012. \label{fig_accel2}}
\end{figure}

Similarly, the first order Fermi acceleration can happen
in terms of the perpendicular momentum.  This is illustrated in
Figure~\ref{fig_accel2}.  There the particle with a large Larmour radius is
bouncing back and forth between converging mirrors of reconnecting magnetic
field systematically getting an increase of the perpendicular component of its
momentum.  Both processes take place in reconnection layers.

\subsection{Force-free turbulent cascade.}
One of the most efficient ways of magnetic dissipation is through the
turbulent cascade where the energy is transported from the large input scales to
the small dissipation scales \citep{Zakharov}.  The non-relativistic theory of
turbulent cascade in magnetized plasma is in the stage of active development,
both theoretical \citep{iro63,kra65,mt81,smm83,sg94,gs97,nb96, lv99, gal00} and
numerical \citep{mg01,clv02, cl02, cl03, kl10}.  The consensus that is emerging from these studies is
that in the non-relativistic case the \Alfven wave cascade is well decoupled
from the fast and slow modes due to the fact that \Alfven waves are incompressible
(slow mode follows \Alfven waves).  Unlike the hydrodynamic turbulence, the \Alfven
wave cascade is anisotropic with energy cascading mostly perpendicularly
to the magnetic field. 

What is the structure of cascade in strongly magnetized plasma? First,
we expect that like the traditional hydromagnetic turbulence the electromagnetic 
turbulence is local in phase space, so that the most important interaction is between waves with 
similar wave lengths. This is because the longer wavelength perturbations can be 
excluded via the relevant Lorentz transformation.  

 Previously, its properties have been 
addressed in \cite{tb98,tp04}. 
\cite{tb98} assumed that the relativistic \Alfven turbulence is similar 
to the non-relativistic one thus the energy cascade remains 
anisotropic.  However, this is unlikely to true for the electromagnetic
turbulence where the fast waves play as an important role as the \Alfven 
waves. Indeed, there exists strong three-wave coupling between 
the non-zero frequency \Alfven waves and fast waves:
$A+F \rightarrow A$, $A+F \rightarrow F$ and $A+A \rightarrow F$. 
In the non-relativistic case, the three wave coupling between non-zero
frequency \Alfven waves and fast waves does not exist because the 
resonance condition cannot be satisfied, and the coupling 
between three fast waves does not exists because of the vanishing coupling 
coefficient.

The three wave interaction coefficients are complicated,  with
different dependences on the angles of interacting waves. 
Generically, the coupling of \Alfven and fast waves is strong,
 so that the two cascades
are well coupled.
Two important questions need to be answered: what are the angular  dependence of the cascades
and what are the wave number dependence.
 It is feasible that both cascades are anisotropic in such a way that
the stationary kinetic equation for both \Alfven and fast waves are satisfied.
We consider this unlikely.
Alternative possibility, which we favor, is that the
  interaction of two cascades may isotropise them.

\section{Step toward relativistic turbulence: vortical  flows of relativistic fluid}

Dynamics of relativistic plasma  is a basic problem in fluid mechanics that has a wide range of applications from the physics of early Universe
 to heavy nuclei collisions to astrophysics. In cosmological applications,   the  post-inflation stage of  reheating, that  lead to matter creation,  is dominated by relativistic turbulence 
 \citep{2003PhRvL..90l1301M}.
  In nuclear physics, the
 head-on collision of two highly relativistic nuclei creates a relativistically  hot  quarkÐgluon plasma that  \citep{1995NuPhA.595..346R}. 
On a very different scale,  a  wide variety of astrophysical objects like jets from Active Galactic Nuclei   \citep{BegelmanBlandfordRees},
Gamma Ray Bursts  \citep{LyutikovJPh}  pulsar winds  \cite{kc84}   contain relativistic plasma. 

Both, the quarkÐgluon plasma  of  nuclear collisions and  astrophysical plasma are nearly ideal,  with very small viscose contribution. Nearly ideal fluids are subject to the development of turbulence \cite{LLVI}. In the astrophysical set-up the  relativistic  turbulence may result in a dynamo action, that may be essential for the production of the high energy emission
\citep{ZhangMacFadyen}. In addition, in case of relativistic supersonic flows the  turbulence is  necessarily for acceleration of cosmic rays at shocks  \cite{BlandfordEichler}. 

Despite these important applications, the theory of relativistic turbulence is not developed, with only a few works addressing its statistical and dynamics properties \citep[\eg][]{1996PhRvE..53.5502D,2008JFM...604..325G}.  Since the relativistic turbulence is generically compressive, we expect
  (following  the analogy with the non-relativistic compressive turbulence) that the turbulence can be represented as a collection of interacting vortical and compressible modes. The compressible modes in the relativistic plasma are well known \citep[\eg][]{LLVI}. On the other hand, the vortical modes in the relativistic fluid has not been considered so far. 

  One of the key features of   relativistic vortices is that the flow compressibility must be taken into account. This make a majority of work on vortices, which often use a non-compressible approximation,  \citep{LLVI,Lamb},  not applicable to  relativistic vortices. On the other hand, relativistic vortices resemble in many ways the vortices in the compressible fluids  \cite{GreenVortex}. The centrifugal forces ``pull away'' gas from the axis  and can lead to the development of a  cavitated core.
\begin{figure}[h!]
\includegraphics[width=0.49\textwidth]{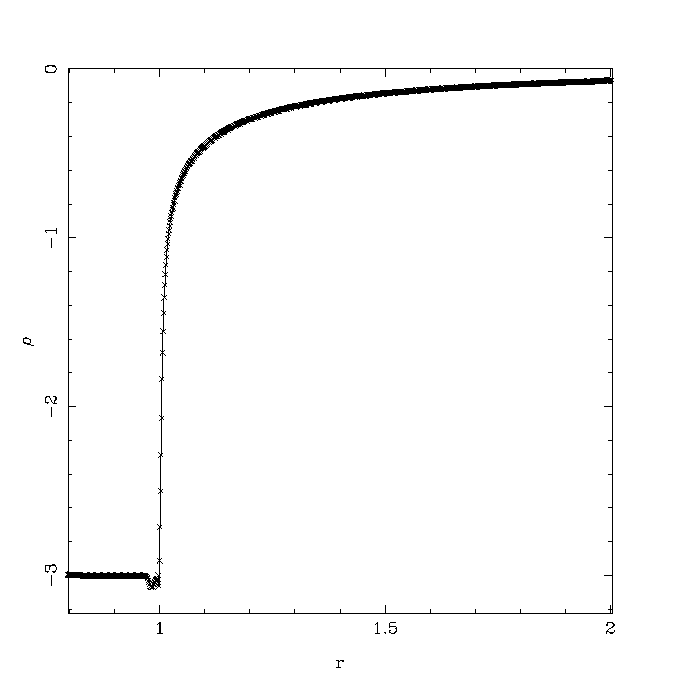}
\includegraphics[width=0.49\textwidth]{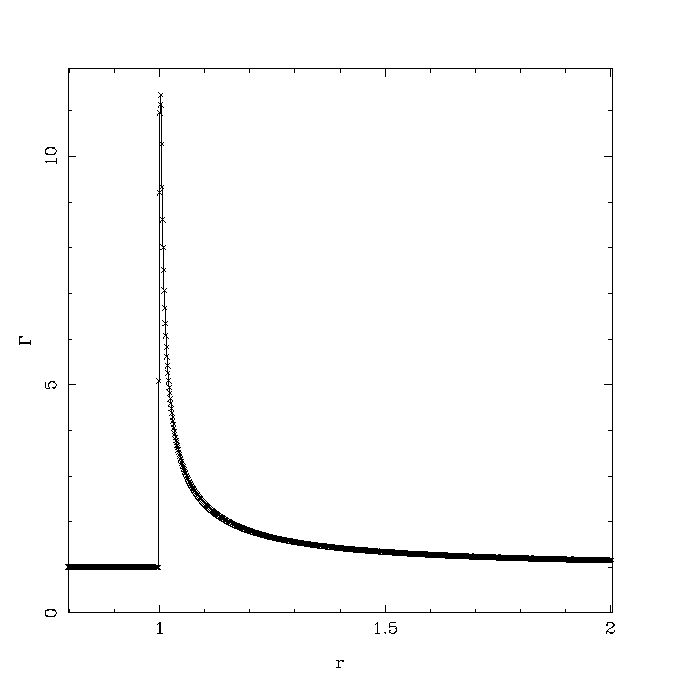}
\caption{ Structure of relativistic vortex: proper density and Lorentz factor. One clearly sees a cavitated core.  The inside of the vortex had to be filled with dynamically unimportant low density gas  since the scheme does not allow full cavitation. }
\label{vortex} 
  \end {figure}
   First we consider a structure of simple rectilinear relativistic vortex. A simple rectilinear vortex with zero distributed vorticity has the following 
 four-velocity structure:
   \be
  u_\phi = {1 \over \sqrt{ (r/r_0)^2 -1}}
  \label{uphi}
  \ee
For polytropic equation of state with index $\Gamma$
the density becomes zero at
\be
{ r_i }  =  { 1+\Gamma \kappa_\infty \over \sqrt{ \Gamma \kappa_\infty (2 + \Gamma \kappa_\infty)}} {\Gamma_z \over 2  \pi \tilde{w }_\infty}  
\ee
where $\kappa_\infty $ and $\tilde{w }_\infty$ are vorticity and proper enthalpy measured at infinity.
Thus, a simple  relativistic vortex is cavitated - it has an empty vacuum core.

Vortical flows must have cores with special properties that  are different from the bulk flow.
The structure of the core of the both relativistic and non-relativistic vortices  depends both on the parameters of the system and the history how a vortex has been created. 
Formally, for realistic  isentropic EoS the  rectilinear relativistic vortex develops a cavitated core.  
The development of the cavitated core occurs in the non-relativistic compressible fluid  as well.  For a  typical  laboratory fluid vortex with  highly subsonic flow  $M\ll 1$ and not   exceptionally high  Reynolds numbers $Re$, the  viscose core forms on time of the order of the dynamical time. On the other hand,  in astrophysics we expect $M\sim 1$, while the Reynolds number  $Re$ is huge. Thus,  in astrophysical conditions  the formation time of a viscose core is very long, so that, generically, an  isolated rectilinear  astrophysical  vortex will be cavitated.

\section{The  subtle role of  global electric fields}

Astrophysical plasmas, typically, do not tolerate large scale electric fields: charge non-neutrality in  highly conducting medium is quickly suppressed by drawing-in of the opposite charges. As a result, the dynamical role of  large-scale electric fields in the overall dynamics and particle acceleration is often under-appreciated. Magnetic fields, on the other hand, may both suppress charge neutralization and can themselves give rise to  global inductive \Efs. Below we discuss two cases where  global inductive \Efs\ and charge non-neutral plasma may be important in otherwise conventional plasmas.

\subsection{Inductive Acceleration of UHERCs}

Relativistic outflows carrying large scale \Bfs have large 
inductive potential and may  accelerate protons to ultra high energies.  \cite{2007APh....27..473L} discussed a  scheme of Ultra-High Energy Cosmic
Ray (UHECR)  acceleration  due to drifts in 
 magnetized, cylindrically collimated, sheared jets of powerful active galaxies.

\cite{2007APh....27..473L}  model of UHECR acceleration relies on the observation that in a
transversely sheared flow one sign of charges is located at a maximum of electric
potential, as we describe in this section. Consider 
 sheared flow carrying \Bf. At each point there is electric field $\E = - {\bf v} \times \B/c$, so that the 
electric potential is determined by
\be
\Delta \Phi = { 1 \over c}  \nabla \cdot \left({\bf v} \times \B  \right)
=   { 1  \over c}  \left( \B \cdot \left( \nabla \times   {\bf v} \right)
-{\bf v} \cdot \left( \nabla \times  \B \right)
\right)
\label{Phii}
\ee
If system is stationary and current-free, 
(in a local rest frame, the second term in Eq. (\ref{Phii})
vanishes at the position of a particle and is generally sub-dominant to the first term in the near
vicinity)
then $\nabla \times  \B=0$ and we find
\be
\Delta \Phi = { 1 \over c} \left( \B \cdot \nabla \times {\bf v} \right)
\label{Phi}
\ee
We have arrived at an important result:
{\it  depending on the sign of the quantity 
$ \left( \B \cdot \nabla \times {\bf v} \right)$ (which is a scalar) charges  of 
one  sign are  near potential minimum, while those with the
opposite sign are near  potential maximum. }
Since electric field is perpendicular both to velocity and magnetic field, locally, the
electric potential is a function of only one coordinate along this direction.
For $ \left( \B \cdot \nabla \times {\bf v} \right) <0 $ 
ions are near potential maximum. 
A
 positively charged particle carried by such a plasma is in an unstable  
equilibrium if  ${\bf B} \cdot \nabla \times {\bf v}< 0$, 
so that kinetic drift along the velocity shear would lead
to fast, {\it regular} energy gain. 

The procedure outlined above to calculate electric potential
is beyond the limits  of applicability of {\it non-relativistic}
MHD, which assumes quasi-neutrality and thus neglects the dynamical effects associated with the 
potential (\ref{Phi}). Thus, even in the low frequency regime with non-relativistic velocities, 
a conventional realm of MHD theory, one should use at least two fluid approach and 
also must retain both charge density as well as
displacement current in Maxwell equations.

Under ideal fluid approximation particles cannot move 
across \Bf lines, so that they cannot
``sample'' the electric potential  (\ref{Phi}). 
On the other hand, kinetic effects, 
like drift motions, may lead to regular radial displacement  along the shear and thus along
electric field.  In this case
 one sign of charge will be gaining energy, while
another sign will be losing energy. This is independent on whether
the drift is along the  shear  or counter to the shear direction
and thus is independent on the sign of the \Bf gradient that induces the shear.

When the Larmor radius becomes comparable to shear scale the 
particle  motion becomes unstable even for
homogeneous  flow.
Particle trajectories
can be found in quadratures in the general case \cite{2007APh....27..473L}. In the  non-relativistic limit
equations of motion can be  integrated exactly,
\be
y= r_L \cos Z \om_B \sqrt{ 1 + \eta /Z \om_B} t, \,
z= - {r_L \over \sqrt{ 1 + \eta /Z \om_B}} \sin Z \om_B \sqrt{ 1 + \eta /\om_B} t
\ee
where $\eta = V'$. 
 This
 clearly shows that for strong negative  shear, $\eta < - \om_B$, particle trajectory is unstable and its energy growth exponentially.
For positive shear, $\eta > 0 $,  particle motion is stable.

\subsection{Relativistic effects at  cosmic ray-modified perpendicular shocks}
Acceleration of \CRs\ is one of the main problems of high energy astrophysics. Shock acceleration is the leading model \citep{BlandfordEichler}. 
Particle acceleration at  quasi-parallel shocks  (when the \Bf\ in the upstream medium is nearly   aligned  with  the shock normal) and  quasi-perpendicular shocks
 (when the \Bf\ in the upstream medium is nearly orthogonal  to the shock normal) proceeds substantially differently. 
 Most astrophysical shocks are quasi-perpendicular, yet  theoretically   acceleration at this type of  shocks is less understood than in the case of quasi-parallel  shocks.
  It is recognized that the feedback of accelerated \CRs\  may considerably modify the parallel shock structure  \citep{1982A&A...111..317A}. 
  \begin{figure}[h!]
\includegraphics[width=.9\linewidth]{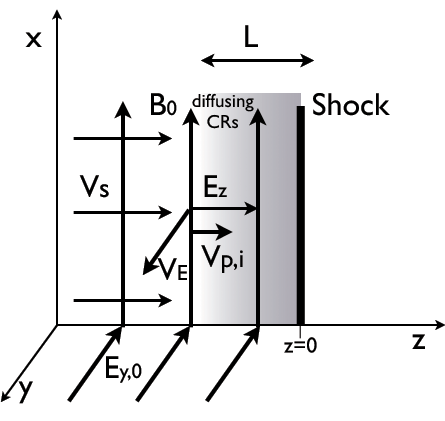}
\caption{Plasma flows in the frame of the shock.
Far upstream, the incoming plasma moves with velocity $v_s$, \Bf\ is along $x$ direction and inductive \Ef\ along $-y$ direction.
At the shock, \CRs\ are accelerated and diffuse ahead of the shock a typical distance $L$, creating an electric field along the shock normal. Electric drift of  \CRs\ in this induced \Ef\   and the  initial \Bf\  produces $z$-dependent electric drift $v_E$  in $y$ direction. Acceleration of plasma in $y$ direction in turn produces polarization drift of ions in $z$ direction.}
\label{perp-picture}
\end{figure}

Kinetic diffusion of cosmic rays ahead of perpendicular shocks induces large scale charge non-neutrality, which is typically neglected in the non-relativistic fluid approach. 
 Cosmic rays  diffusing ahead of the shock offset the charge balance in the incoming plasma, which becomes non-neutral, with \Ef\ directed along the shock normal. The incoming upstream plasma will {\it partially } compensate this charge density by a combination of electric and polarization drifts. This creates a current along the shock normal, perpendicular to the \Bf, see Fig. \ref{perp-picture}.  Current-driven instabilities, in particular of the modified Buneman type,  generate plasma turbulence with  wave vectors preferentially perpendicular to the initial \Bf, generating the field line wandering required for  acceleration of \CRs\ in the first place. Thus, similar to parallel shocks, assumption of turbulence and \CR\ acceleration leads to turbulence generation by \CRs\ themselves \cite{2010MNRAS.407.1721L}.

  \newpage
  
\bibliographystyle{apj}

\end{document}